# Time in Quantum Mechanics and Quantum Field Theory

Z. Y. Wang[1, 2, *] and B. Chen[3]

[1]*P. O. Box 1, Xindu, Chengdu, Sichuan, 610500, P. R. China*

[2]*00C01003, University of Electronic Science and technology of China, Chengdu, Sichuan 610054, P. R. China*

[3]*School of Optics/CREOL, University of Central Florida, Orlando, FL 32816, USA*

## Abstract

W. Pauli pointed out that the existence of a self-adjoint time operator is incompatible with the semibounded character of the Hamiltonian spectrum. As a result, people have been arguing a lot about the time-energy uncertainty relation and other related issues. In this article, we show in details that Pauli's definition of time operator is erroneous in several respects. In order to define time operator correctly, by treating time and space on an equal footing and extending the usual Hamiltonian $\hat{H}$ to the generalized Hamiltonian $\hat{H}_\mu$ (with $\hat{H}_0 = \hat{H}$), we reconstruct the analytical mechanics and the corresponding quantum (field) theories, which are equivalent to the traditional ones. The generalized Schrödinger equation $i\partial_\mu \psi = \hat{H}_\mu \psi$ and Heisenberg equation $d\hat{F}/dx^\mu = \partial_\mu \hat{F} + i[\hat{H}_\mu, \hat{F}]$ are obtained, from which we have: 1) $t$ is to $\hat{H}_0$ as $x_j$ is to $\hat{H}_j$ ($j=1,2,3$); likewise, $t$ is to $i\partial_0$ as $x_j$ is to $i\partial_j$; 2) the correct time operator is canonical conjugate to $i\partial_0$ rather than $\hat{H}_0$, the Pauli's theorem no longer holds; 3) there have two types of uncertainty relations: the usual $\Delta x_\mu \Delta p_\mu \geq 1/2$ and the Mandelstam-Tamm's treatment $\Delta x_\mu \Delta H_\mu \geq 1/2$.

[*]Electronic address: wangzhiyong168@yahoo.com.cn

# 1. Introduction

Time in quantum mechanics has been a controversial issue since the advent of quantum theory. Nowadays it still has theoretical and practical interests.

On the one hand, there exist enough reasons for us to consider time as a dynamical variable or operator: 1). According to the relativity, a position vector *operator* in a reference system would have a temporal component in another reference system; 2). The 4-dimensional (4D) angular momentum tensor operator of quantum field show that, in the quantum field theory, time seems to play a twofold role, as a parameter and an operator; 3) A major conceptual problem in quantum gravity is the issue of what time is, and how it has to be treated in the formalism;[1] 4). In many cases, time is not a mere parameter, but an intrinsic property characterizing the duration of certain physical processes. The lifetime of unstable particles or collision complexes is a well-known example; 5). Another related problem, which still remains controversial today, is concerned with the formal definitions of traversal and tunneling time.[2-11] This subtle question, motivated in part by the possible applications of tunneling in semiconductor technology, has received considerable attention in recent years; 6) In signal analysis and signal processing theory,[12] as a physical quantity, time and frequency are treated on an complete equal footing; 7). Lack of an appropriate time operator has a number of consequences. In particular, the time-energy uncertainty relation has remained ambiguous over these years and its improper application has led to a great deal of confusion.[13-19]

But on the other hand, as well known, according to Pauli's argument,[20] the existence of a self-adjoint time operator is incompatible with the semibounded character of the Hamiltonian spectrum. By using a different argument based on the time-translation property of the arrival time concept, Allcock has found the same negative conclusion.[21] This negative conclusion can be also traced back to the semi-infinite nature of the Hamiltonian spectrum.

Then, we are in a dilemma, but what is wrong? In this article, we try to solve this problem.   In this work we apply natural units of measurement ($\hbar = c = 1$).

## 2. The flaws of Pauli's statement

According to Pauli's statement, let *time operator* $\hat{T}$ satisfy $[\hat{H},\hat{T}] = -i$, we have $[f,\hat{H}] = i\frac{\partial f}{\partial t}$; let $\hat{H}\psi_E = E\psi_E$, and then we have $\hat{H}e^{i\alpha\hat{T}}\psi_E = (E+\alpha)e^{i\alpha\hat{T}}\psi_E$, where $\alpha$ is an arbitrary constant.   That is, $e^{i\alpha\hat{T}}\psi_E$ is also the eigenstate of the Hamiltonian $\hat{H}$ with the eigenvalue ($E+\alpha$), which implies that the existence of time operator contradicts the fact that the Hamiltonian spectrum must be positive.

However, Pauli's (or Allcock's) statement is erroneous in three respects, which is owing to the incorrect definition of time operator:

(1). Pauli's (or Allcock's) demonstration implies a premise that the *time* operator itself is not explicitly *time* dependent: $\frac{d\hat{T}}{dt} = \frac{\partial \hat{T}}{\partial t} + i[\hat{H},\hat{T}] = i[\hat{H},\hat{T}]$.   However, studying the conservative property of the 4D angular momentum tensor of a free field (e.g., the Dirac field), we find that, in contradiction to Pauli's (or Allcock's) statement, we have $\frac{d\hat{T}}{dt} = \frac{\partial \hat{T}}{\partial t} + i[\hat{H},\hat{T}] = \frac{\partial \hat{T}}{\partial t}$ (see Appendix A).   That is to say, in Heisenberg's picture, the time operator is explicitly time dependent, just as that the position operator is explicitly dependent of position coordinate.

In fact, the 4D angular momentum tensor of a charged field is related to the electromagnetic moment tensor (see Appendix B), and has observable effects.[22]

(2). As will be shown later, time operator defined correctly is canonical conjugate to $i\frac{\partial}{\partial t}$ rather than $\hat{H}$, correspondingly, Pauli's (or Allcock's) statement no longer holds: this moment it is just a matter of anew choosing for zero-energy reference surface.

(3). In logic, a self-related proposition can lead to a paradox.   Likewise, if A in the equation of motion $\frac{dA}{dt} = f$ is time itself, the traditional method that translates the classical formulation into the quantum framework no longer holds true, and all the related theory must be reformulated in a different approach.

In fact, someone has shown that Pauli's implicit assumptions are not consistent in a single Hilbert space.[23]

### 3. The starting point for introducing time operator

As mentioned above, an important motive for trying to introduce time operator lies in that, the theory of relativity requires that time and space must be treated on an equal footing, which means that a premise for introducing time operator is putting time coordinate on the same footing as position coordinates. However, the traditional theories treat time and space very differently:

1) The traditional many-particle system theory has a defect: the system contains only one time variable while there are as many position variables as there are particles, which in contradiction with the relative simultaneity: observing in another reference system, all particles in the system could not share a common time coordinate, and the original Hamiltonian no longer corresponds to *the total energy* that stands for the energy sum of all particles at the same moment of time (note that the energy of single particle in the system is not necessary conservative). Historically, in view of this unsatisfactory aspect of the traditional theory, people have introduced the many-time formalism theory[24] (being equivalent to the Heisenberg-Pauli theory), where for a system composed of $N$ particles there correspond to N distinct time and space variables. In this sense, time and space are treated with equality[25] in the many-time formalism.

2) However, even in the many-time formalism of quantum mechanics, for every single particle of a many-particle system, its time and space coordinates are still not equal, namely, its space coordinates can be taken as dynamics variables while time coordinate not, this is what we will try to solve in this paper. In view of mentioned above, our discussion would only be limited to the relativistic single particle and quantum field cases.

Certainly, even in the theory of relativity, time and space could not be complete equal because of the law of causality. In other words, in space-time diagrams, the distribution of the worldline of an arbitrary particle is not symmetrical about the surface of lightcone.

In a word, in order to define time operator correctly, it is necessary to put time coordinate on the same footing as position coordinates. For this we reconstruct the analytical mechanics and the corresponding quantum theories, which are equivalent to the traditional ones.

### 3.1 The generalized Schrödinger equation

Starting from the usual relativistic quantum mechanics equations, we can obtain the generalized Schrödinger equations as follows:

$$i\partial_\mu \psi(x) = \hat{H}_\mu \psi(x) \quad (\mu = 0,1,2,3) \tag{1}$$

where $\hat{H}_\mu$ is the generalized Hamiltonian with $\hat{H}_0 = \hat{H}$ being the usual Hamiltonian.

#### 3.1.1 The generalized Schrödinger form of the Klein-Gordon equation

As for the Klein-Gordon equation

$$(\partial^\mu \partial_\mu + m^2)\phi = 0 \tag{2}$$

people have found a Schrödinger formulation[26] of Eq.2, where Eq.2 is transformed into Eq.1 with $\mu = 0$. Now, we will show that Eq.2 can also read as Eq.1 with $\mu = l = 1,2,3$. Let (Fortunately, the massless spin-0 particle does not exist)

$$\begin{aligned}\varphi &= \frac{1}{2}(i - \frac{i}{m}\frac{\partial}{\partial x^l})\phi \\ \chi &= \frac{1}{2}(i + \frac{i}{m}\frac{\partial}{\partial x^l})\phi\end{aligned} \quad (l=1,2,3) \tag{3}$$

$$\psi = \begin{pmatrix} \varphi \\ \chi \end{pmatrix}, \quad \tau_2 = \begin{pmatrix} 0 & -i \\ i & 0 \end{pmatrix}, \quad \tau_3 = \begin{pmatrix} 1 & 0 \\ 0 & -1 \end{pmatrix}. \tag{4}$$

Using Eqs.3 and 4, we can express Eq.2 as Eq.1 with $\mu = l = 1,2,3$, where

$$\hat{H}_l = -\frac{i}{2m}(\tau_3 + i\tau_2)(\partial^\nu \partial_\nu - \partial^l \partial_l) - im\tau_3 \tag{5}$$

note that the repeated indices in $\partial^\nu \partial_\nu$ is summed while in $\partial^l \partial_l$ not, $\nu = 0,1,2,3$ while $l$ is one of $1,2,3$. For the time being, the scalar product and the expectation value are defined as

$$\langle\psi|\psi'\rangle \equiv \int d\sigma^l \psi^+ \tau_2 \psi$$
$$\langle\hat{L}\rangle \equiv \int d\sigma^l \psi^+ \tau_2 \hat{L}\psi \quad (l=1,2,3) \tag{6}$$

respectively, where $d\sigma^\mu \equiv (dx^1 dx^2 dx^3, dt dx^2 dx^3, dx^1 dt dx^3, dx^1 dx^2 dt) = (d\sigma^0, d\vec{\sigma})$ stands for a 3D hypersurface element.

### 3.1.2 The generalized Schrödinger form of the Dirac equation

Substitute

$$\hat{H}_\mu = -\gamma_\mu (i\gamma^\nu \partial_\nu - i\gamma^\mu \partial_\mu) + \gamma_\mu m \quad (\mu = 0,1,2,3) \tag{7}$$

into Eq.1, we can obtain the same Dirac equation for $\mu = 0,1,2,3$, where $\gamma^\mu$ are the Dirac matrices, the repeated indices in $i\gamma^\nu \partial_\nu$ is summed while in $i\gamma^\mu \partial_\mu$ not. In contrast to the case of the Klein-Gordon equation, here the scalar product and the expectation value are defined as the traditional ones.

All mentioned above also hold in the presence of interactions. For example, one just makes the replacement $\partial_\mu \to \partial_\mu + ieA_\mu$ for each equation above that contains $\partial_\mu$ ($\mu = 0,1,2,3$).

### 3.2 The generalized Heisenberg's equation

Using Eq.1 and the corresponding definitions of expectation value, one can obtain

$$\frac{d<\hat{F}>}{dx^\mu} = <\frac{\partial \hat{F}}{\partial x^\mu}> + i<[\hat{H}_\mu, \hat{F}]> \tag{8}$$

where $<\hat{F}>$ is the expectation value of a dynamical operator $\hat{F}$.

In general, by enlarging $t$ to $x^\mu$, we can arrive at the 4D generalization of the time-evolution operator (say, the space-time evolution operator). Furthermore, in an analogous procedure we can arrive at the four-dimensional generalization of Heisenberg's equation

$$\frac{d\hat{F}}{dx^\mu} = \frac{\partial \hat{F}}{\partial x^\mu} + i[\hat{H}_\mu, \hat{F}] \tag{9}$$

Traditionally, the four-dimensional generalization of Heisenberg's equation is[27]

$$\frac{\partial \hat{F}}{\partial x^\mu} = i[\hat{p}_\mu, \hat{F}] \tag{10}$$

However, from Eq.10 we can only obtain $\frac{\partial \hat{F}}{\partial x^\mu} = \frac{\partial \hat{F}}{\partial x^\mu}$. That is to say, in contrast to Eq.9, Eq.10 is only a mathematical identity (without any physical content) rather than a physical equation.

In addition, using Eq.9, one can obtain

$$\Delta x_\mu \Delta H_\mu \geq \frac{1}{2} \tag{11}$$

where $\Delta H_\mu \equiv \sqrt{\langle (\hat{H} - \langle \hat{H} \rangle)^2 \rangle}$ is the uncertainty (the mean variation) of $\hat{H}$, and $\Delta x_\mu$ is defined by $\Delta x_\mu \left| \left\langle \frac{d\hat{F}}{dx^\mu} \right\rangle \right| = \Delta F$. That is to say, the Mandelstam-Tamm's treatment[28] of the time-energy uncertainty relation (as it was formulated in most textbooks) has a counterpart of the position-momentum uncertainty relation, which implies that, as for the same mathematical expression, it can carry a variety of physical content and hence can exist a variety of physical interpretation.

### 3.3 The generalization in quantum field theory

Before going on, let us introduce the generalized analytical mechanics. Let $q$ be the generalized coordinates of a system, p the generalized momentums, $L$ the Lagrangian, $H$ the Hamiltonian and $A$ the action. *Generalized analytical mechanics* can be obtained by extending $t$ to $x^\mu$ (let $\mu = 1$ without loss of generality). For this we make the following replacements:

$$A = \int L dt \rightarrow A = \int L^1 dx^1 \quad \text{(From which we define } L^1\text{)} \tag{12}$$

$$q(t) \rightarrow q(x^1) \ , \ L(q(t), \dot{q}(t)) \rightarrow L^1(q(x^1), \partial_1 q(x^1)) \quad \text{(By definition)} \tag{13}$$

Using Eqs.12 and 13, we have

$$\frac{\partial L}{\partial q} - \frac{d}{dt}\frac{\partial L}{\partial \dot{q}} = 0 \rightarrow \frac{\partial L^1}{\partial q} - \partial^1 \frac{\partial L^1}{\partial \partial^1 q} = 0 \quad \text{(By the principle of least action)} \tag{14}$$

In fact, Eq.14 is a special case for the Whittaker equation.[29] Furthermore

$$p \equiv \frac{\partial L}{\partial \dot{q}} \to p_{(1)} \equiv \frac{\partial L^1}{\partial \partial^1 q} \quad \text{(By definition)} \tag{15}$$

$$\dot{p} = \frac{\partial L}{\partial q} \to \partial^1 p_{(1)} = \frac{\partial L^1}{\partial q} \quad \text{(By Eqs.14 and 15)} \tag{16}$$

$$H \equiv p\dot{q} - L \to H_1 \equiv p_{(1)} \partial_1 q - L_1 \quad \text{(By definition)} \tag{17}$$

$$\dot{q} = \frac{\partial H}{\partial p}, \dot{p} = -\frac{\partial H}{\partial q} \to \partial_1 q = \frac{\partial H_1}{\partial p_{(1)}}, \partial_1 p_{(1)} = -\frac{\partial H_1}{\partial q} \quad \text{(By Eqs.12-17)} \tag{18}$$

$$\frac{df}{dt} = \frac{\partial f}{\partial t} + \{H, f\} \to \frac{df}{dx^1} = \frac{\partial f}{\partial x^1} + \{H_1, f\} \quad \text{(By Eqs.18)} \tag{19}$$

Where $f = f(q, p_{(1)}, x^1)$ and $\{H_1, f\} = \frac{\partial f}{\partial q}\frac{\partial H_1}{\partial p_{(1)}} - \frac{\partial f}{\partial p_{(1)}}\frac{\partial H_1}{\partial q}$.

Clearly, all mentioned here are based on the first principles and do not resort to any heuristic argument, by which we put $t$ on the same footing as $\bar{x}$ in our analytical mechanics. The correctness of such formalism can be further shown later.

The generalization mentioned above is also valid for quantum field theory. Let $d\sigma^\mu$ ($\mu = 0,1,2,3$) stand for a 3D hypersurface element, if the action $A = \int d^4 x\, \Gamma$, where $\Gamma$ stands for the Lagrange density, using Eqs. 12, we have

$$L^\mu = \int d\sigma^\mu \Gamma \qquad (\mu = 0,1,2,3) \tag{20}$$

Obviously, $L^0 = L$ is the usual Lagrangian.

We assume that the 3D hypersurface $\sigma^l$ ($l = 0,1,2,3$) is divided into small cells of size $\Delta\sigma_i^l$. With each cell we associate the respective average values of the functions, e.g.:

$$\phi_i(x^l) = \frac{1}{\Delta\sigma_i^l} \int_{\Delta\sigma_i^l} \phi(\bar{x}, t) d\sigma^l \tag{21}$$

As $\Delta\sigma_i^l \to 0$, $\phi_i(x^l) \to \phi(\bar{x}, t) \equiv \phi(x)$. By applying Eqs.12-19 and the proceedings analogous to the traditional process of canonical quantization, we can obtain the following results: the generalized canonically conjugate field of $\phi(\bar{x}, t)$ is defined as

$$\pi_l(\bar{x}, t) \equiv \frac{\partial \Gamma(x)}{\partial \partial^l \phi(x)} \qquad (l = 0,1,2,3) \tag{22}$$

and the generalized Hamiltonian is

$$H_1(x^1) \equiv \int [\pi_1(x)\partial_1\phi(x) - g_{ll}\Gamma(x)]d\sigma^1 \qquad (23)$$

where $g_{\mu\nu}$ is the metric tensor with $\mathrm{diag}(1,-1,-1,-1)$, the repeated index $l$ is not summed. Obviously, $T_{\mu\nu} = \pi_\mu \partial_\nu \phi - g_{\mu\nu}\Gamma$ is the energy-momentum tensor of a field (see Appendix C).

In the case of quantum field theory, however, the condition of microcausality must be taken into account. Traditionally, we firstly study the plane-wave solutions of a free field equation and obtain $p_0^2 = p_1^2 + p_2^2 + p_3^2 + m^2$ (in general, we can call $w_\mu \equiv \sqrt{p_\mu^2}$ ($\mu = 0,1,2,3$) as *frequency* or *wave number in* $x^\mu$), and then write the general solution as a linear combination of the $\pm w_0$ solutions. The general solution contains the factors $e^{\pm ip\cdot x}$, where $p\cdot x = w_0 t - p_1 x_1 - p_2 x_2 - p_3 x_3$, $w_0 \geq 0$ while $p_1, p_2, p_3 \in (-\infty,+\infty)$. Now, in our case (make the replacement $t \to x^1$ without loss of generality), in order to preserve microcausality, when we obtain $p_0^2 = p_1^2 + p_2^2 + p_3^2 + m^2$, i.e. $p_1^2 = p_0^2 - p_2^2 - p_3^2 - m^2$, we rewrite the general solution as a linear combination of the $\pm w_1 \equiv \pm\sqrt{p_1^2}$ solutions. This moment in the factors $e^{\pm ip\cdot x}$, $p\cdot x = p_0 t - w_1 x_1 - p_2 x_2 - p_3 x_3$, where $w_1 \equiv \sqrt{p_1^2} \geq 0$ while $p_0, p_2, p_3 \in (-\infty,+\infty)$.

In the following we will take the Klein-Gordon field and the Dirac field for example, while the photon field is analogous to the former.

### 3.3.1 The Klein-Gordon field

The Lagrange density of the Klein-Gordon field $\phi(\bar{x},t) = \phi(x)$ reads

$$\Gamma(x) = \partial_\mu \phi^+(x)\partial^\mu \phi(x) - m^2 \phi^+(x)\phi(x) \qquad (24)$$

where $\phi^+$ is the Hermitian adjoint of $\phi$. Let $w_\mu \equiv \sqrt{k_\mu^2}$ and $k\cdot x = k_l x^l + w_\mu x^\mu$ (the repeated index $\mu$ is not summed), where $w_\mu \geq 0$ and $k_l \in (-\infty,+\infty)$, $\mu \neq l = 0,1,2,3$. One can easily show that the fields $\phi$ and $\phi^+$ can also be expressed as a linear combination of the $\pm w_\mu$ solutions

$$\phi(x) = \int d\sigma_k^\mu (a_k u_k(x) + b_k^+ u_k^*) \qquad (25)$$

where $u_k(x) = [2w_1(2\pi)^3]^{-\frac{1}{2}} e^{-ik \cdot x}$, $d\sigma_k^\mu$ is the $\mu$-component of the 3D hypersurface element in 4-momentum space. As $w_\mu \to 0$, all our final results (in the observable sense) also hold. By defining $a\overset{\leftrightarrow}{\partial} b \equiv a\partial_\mu b - (\partial_\mu a)b$, we have

$$a_k = \int d\sigma^\mu u_k^*(x) i\overset{\leftrightarrow}{\partial} \phi(x)$$

$$\int d\sigma^\mu u_{k'}^*(x) i\overset{\leftrightarrow}{\partial} u_k(x) = \delta^3(k_j - k'_j) \quad (\mu \neq j = 0,1,2,3) \tag{26}$$

and so on. Using Eq.22 we have

$$\pi_\mu(x) = \partial_\mu \phi^+(x), \pi_\mu^+ = \partial_\mu \phi(x) \tag{27}$$

In view of the fact that, what we finally utilize is the canonical conjugate commentators rather than the so-called covariant commentators (the former correspond to the derivative of the latter), and we only discuss the former. Generally, when $\phi$ and $\phi^+$ are expressed as a linear combination of the $\pm w_\mu (w_\mu \equiv \sqrt{k_\mu^2})$ solutions, we have

$$[\phi(x), \pi_\mu(y)] = i\int d^4k [w_\mu \delta(k^2 - m^2) e^{-ik \cdot (x-y)}] \tag{28}$$

$$[\phi(x), \pi_\mu(y)]\big|_{x_\mu \to y_\mu} = i\delta^3(x^l - y^l) \quad (\mu, l = 0,1,2,3, \mu \neq l) \tag{29}$$

and so on.. Using Eq.23, we have

$$H_\mu = \int d\sigma^\mu [\pi_\mu \partial_\mu \phi + \pi_\mu^+ \partial_\mu \phi^+ + \Gamma]$$
$$= \int \frac{d\sigma_k^\mu}{(2\pi)^3} w_\mu(k)[a^+(k)a(k) + b^+(k)b(k) + 1] \tag{30}$$

In view of $w_\mu = w_\mu(k) \geq 0$ (but not $w_\mu(k) \equiv 0$ for arbitrary $k_l, \mu, l = 0,1,2,3, \mu \neq l$), the generalized Hamiltonians $H_\mu$ are always positive for the Bose field. On the one hand, $\phi(x)$ is written as a Hilbert space operator, which creates and destroys the particles that are the quanta of field excitation. On the other hand, $\phi(x)$ is written as a linear combination of the $\pm w_\mu$ ($\mu$ is one of 0,1,2,3) solutions of the Klein-Gordon equation. Both signs of the $x^\mu$-dependence in the exponential appear, although $w_\mu$ is always positive (as mentioned before, our final results hold also for $w_\mu \to 0$). If $\phi(x)$ is single-particle wavefunction, it would correspond to states of positive- and negative-frequency ($\pm w_\mu$) modes. The connection between the particle creation operators and the waveforms displayed here is always valid for free quantum fields: A negative-frequency solution of the field equation, being the Hermitian conjugate of a positive-frequency solution, has as its coefficient the operator that creates a particle in that positive-frequency single-particle wavefunction. In this way, the fact that the related equations have both positive- and negative-frequency solutions (because of $k_0^2 = k_1^2 + k_2^2 + k_3^2 + m^2$ being always valid) is reconciled with the requirement that a sensible quantum theory contain only positive generalized Hamiltonians.

For the photon field, the zero-point contribution to the generalized Hamiltonians $H_\mu$ may lead to *generalized* Casimir effects,[30] which may be verified by a different experimental set-up for a different $\mu = 0,1,2,3$ (for $\mu = 1,2,3$, the Casimir force is related to the time-varying difference $\Delta H_\mu$, which will be discussed in our next paper).

Finally, the Heisenberg's equations of motion are

$$\partial_\mu \phi = i[H_\mu, \phi] \tag{31}$$

$$\partial_\mu \pi_\mu = i[H_\mu, \pi_\mu], \tag{32}$$

and so on, from which the Klein-Gordon equations can be obtained.

**3.3.2 The Dirac field**

As for the Dirac field equation, however, its $\pm w_l$ ($l = 1,2,3$) solutions are not orthogonal. Then, our discussions should be given in another manner. According to the traditional procedure of transforming the quantum mechanics description into the quantum field theory one, our discussions can be carried out on the basis of Eqs.1 and 7. For this, we reinterpret $\psi(x)$ in Eq.1 as field operator that obey the canonical anticommutation rules:

$$\{\psi_\alpha(\bar{x},t), \psi_\beta^+(\bar{x}',t)\} = \delta_{\alpha\beta} \delta^3(\bar{x} - \bar{x}')$$
$$\{\psi_\alpha(\bar{x},t), \psi_\beta(\bar{x}',t)\} = \{\psi_\alpha^+(\bar{x},t), \psi_\beta^+(\bar{x}',t)\} = 0 \tag{33}$$

the generalized Hamiltonian is

$$H_\mu = \int \psi^+(x)\hat{H}_\mu\psi(x)d^3x \tag{34}$$

where $\hat{H}_\mu$ is given by Eq.7. The dynamics of the field operators is determined by the generalized Heisenberg's equations of motion

$$\frac{\partial \psi(x)}{\partial x^\mu} = i[H_\mu, \psi(x)] \tag{35}$$

$$\frac{\partial \psi^+(x)}{\partial x^\mu} = i[H_\mu, \psi^+(x)] \tag{36}$$

e.g., with the help of Eq.33 one finds that Eq.35 leads back to Eq.1 (and hence the Dirac equation). Furthermore, using Eq.34 we have

$$H_\mu = \int d^3p \sum_s p_\mu [c^+(p,s)c(p,s) + d^+(p,s)d(p,s) + \frac{1}{2}\delta_{\mu 0}] \tag{37}$$

where $c^+, c$ (or $d^+, d$) are the creator and annihilator of a particle (or an antiparticle), respectively. $p_\mu$ is the 4-momentum of a single Fourier mode of the field. Obviously, $H_\mu$ is the total 4-momentum of the field, which implies that

$$H_\mu = P_\mu = \int \psi^+(x)\hat{P}_\mu\psi(x)d^3x \quad (\mu = 0,1,2,3) \tag{38}$$

where $\hat{P}_\mu = i\frac{\partial}{\partial x^\mu}$. However, in spite of Eq.38, if we rewrite Eq.35 as

$$\frac{\partial \psi(x)}{\partial x^\mu} = i[P_\mu, \psi(x)] \tag{39}$$

in contrary to Eq.35, Eq.39 gives $\frac{\partial \psi(x)}{\partial x^\mu} = \frac{\partial \psi(x)}{\partial x^\mu}$ instead of leading back to Eq.1.

### 3.3.3 Interacting Quantum Fields

In the following, we will take quantum electrodynamics (QED) for example, where the Hamilton density describing the interaction is given by

$$H_{int} = e\bar{\psi}(x)\gamma^\mu\psi(x)A_\mu(x) \tag{40}$$

In the interaction picture, the field operators $\psi(x)$ and $A_\mu(x)$ are the same as the ones of the free fields. Meanwhile, in our formalism, the electromagnetic vector potential $A_\mu(x)$ is written as a linear combination of the $\pm w_1 (w_1 \equiv \sqrt{k_1^2})$ solutions

$$A_\mu(x) = \frac{1}{\sqrt{(2\pi)^3}} \int d\sigma_k^1 \frac{1}{\sqrt{2w_1}} \sum_\lambda e_\mu(k,\lambda)(a_{k\lambda} e^{-ik\cdot x} + a_{k\lambda}^+ e^{ik\cdot x}) \qquad (41)$$

where $e_\mu(k,\lambda)$ ($\mu = 0,1,2,3$) are the polarization 4-vectors, $\lambda = 0,1,2,3$ the polarization indices, $k\cdot x = k_0 t - w_1 x_1 - k_2 x_2 - k_3 x_3$, $w_1 \equiv \sqrt{k_1^2}$ while $k_0, k_1, k_3 \in (-\infty, +\infty)$. As mentioned before, our final results also hold for $w_1 \to 0$.

To apply Wick's theorem, as for $A_\mu(x)$, we must generalize the definitions of the time-ordering product and the time-evolution operator $x^\mu$ to the $x^\mu$-*ordering product* and $x^\mu$-*evolution operator*, respectively. e.g., the $x^1$-ordering product is

$$T_1 A_\mu(x) A_\nu(y) \equiv \begin{cases} A_\mu(x) A_\nu(y), & x^1 > y^1 \\ A_\mu(y) A_\nu(x), & x^1 < y^1 \end{cases} \qquad (42)$$

and $x^1$-evolution operator $U(x_1, x_1')$ is defined by

$$|a(x_1)\rangle = U(x_1, x_1')|a(x_1')\rangle \qquad (43)$$

From Eqs. 41 and 42, it can be shown that the $x^1$-*Feynman propagator* for photons is the same as the usual $t$-Feynman propagator:

$$\langle 0|T_1 A_\mu(x) A_\nu(y)|0\rangle = \frac{1}{(2\pi)^4} \int d^4 k \frac{-ig_{\mu\nu}}{k^2 + i\varepsilon} e^{-ik\cdot(x-y)} \qquad (44)$$

Furthermore, in the interaction picture, according to our formalism, we have

$$i\frac{\partial}{\partial x^1}|a\rangle = H_{int}|a\rangle \qquad (45)$$

Let $S \equiv U(+\infty, -\infty)$, from Eqs.43 and 45, we have

$$S = T_1 \exp[-i\int d^4 x H_{int}(x)] \qquad (46)$$

We define the contraction of $A_\mu(x)$ in the traditional way with all the related definitions for the Dirac field $\psi(x)$ are the same as the traditional ones except for replacing the time-ordering product with the $x^1$-ordering product. Correspondingly, in Eq.46, we apply the $x^1$-ordering product and the corresponding contraction for $A_\mu(x)$ while keep the same as usual time-ordering product and contraction for $\psi(x)$ (choosing a frame of reference in which $t_i > t_j$ as $x_i^1 > x_j^1$). In this way, the Feynman rules can be obtained, from which we can perform some real calculations, such as the particle-scattering process, where the initial state particles come from $(t, x^1) = (-\infty, -\infty)$ and the final state particles go to $(t, x^1) = (+\infty, +\infty)$, the results are the same as the usual ones.

## 4. Time Operator

Up to now, we treat time and space on an equal footing by extending the usual Hamiltonian $\hat{H}$ to the generalized Hamiltonian $\hat{H}_\mu$ (with $\hat{H}_0 = \hat{H}$), which will provide the basis for us to discuss time operator correctly. Firstly, let us refer to the following facts:

1) From the generalized analytical mechanics to quantum field theory, $t$ is to

    $\hat{H}_0 = \hat{H}$ as $x_j$ is to $\hat{H}_j$ ($j = 1,2,3$); likewise, $t$ is to $i\dfrac{\partial}{\partial t}$ as $x_j$ is to $i\dfrac{\partial}{\partial x_j}$.

2) $\hat{H}_\mu$ is not identically equal to $\hat{p}_\mu$, otherwise if $\hat{H}_\mu \equiv i\partial_\mu$, Eq.1 becomes a purely mathematical identity $\hat{H}_\mu \psi \equiv i\partial_\mu \psi$ (where $\psi$ can be arbitrary), no longer a physical equation.

3) In spite of 2), owing to Eq.1, $i\partial_\mu$ and $\hat{H}_\mu$ have the same spectrum distributions in the same Hilbert space.

4) Translating the classical mechanics into the relativistic quantum mechanics, we make the replacement $p_\mu \to \hat{p}_\mu \equiv i\partial_\mu$ rather than $p_\mu \to \hat{H}_\mu$ ($\mu = 0,1,2,3$).

5) The generator of translation in $x_\mu$ is directly $\hat{p}_\mu$ rather than $\hat{H}_\mu$, only make use of Eq.1 can one also express the generator as $\hat{H}_\mu$.

6) Sometimes, one finds that $[\hat{H}_\mu, x_\mu] = 0$ (by Eqs.5 and 7 for example), whereas $[\hat{p}_\mu, x_\mu] \equiv ig_{\mu\nu}$.

7) In the 4D angular momentum tensor $J^{\mu\nu} = ix^\mu \partial^\nu - ix^\nu \partial^\mu$, $x_\mu$ ($\mu = 0,1,2,3$) is taken as dynamics operator.

From 1)-7), we can draw a conclusion as follows: (1) the defining expression for 4-momentum operator $\hat{p}_\mu$ is $i\partial_\mu$, and its form is the same for all fields; while $\hat{H}_\mu$, as the calculating expression for 4-momentum $\hat{p}_\mu$ (in quantum field theory case, it is $\sqrt{\hat{p}_\mu^2}$ instead of $\hat{p}_\mu$ for the Bose fields), has a different form for a different field; (2) the 4-vector $x_\mu$ is canonically conjugate to the 4-vector $\hat{p}_\mu = i\partial_\mu$ rather than $\hat{H}_\mu$; (3) $t$ can play a twofold roles, as a parameter or a dynamics variable.

In other words, just as Dirac thought,[31] in the position space, the time operator $\hat{T}$ is $t$ itself, but in contrary to his viewpoint, the time operator $t$ is canonically conjugate to $\hat{p}_0 = i\dfrac{\partial}{\partial t}$ rather than the Hamiltonian $\hat{H}_0 = \hat{H}$. $i\dfrac{\partial}{\partial t}$ is the defining expression for energy operator while $\hat{H}$ corresponds to the calculating expression.

As a consequence, one can readily verify that the existence of a self-adjoint time operator (i.e. $t$) is NOT incompatible with the semibounded character of the Hamiltonian spectrum.

In fact, let $i\dfrac{\partial}{\partial t}\psi_E = \hat{H}\psi_E = E\psi_E$, similar to Pauli's argument, using $[f, i\dfrac{\partial}{\partial t}] = -i\dfrac{\partial f}{\partial t}$, we have

$$i\dfrac{\partial}{\partial t}e^{i\alpha t}\psi_E = (E-\alpha)e^{i\alpha t}\psi_E, \text{ i.e. } (i\dfrac{\partial}{\partial t} - \alpha)\psi_E = (\hat{H} - \alpha)\psi_E = \hat{H}'\psi_E = (E-\alpha)\psi_E.$$

Therefore, it is just a matter of the choice of zero-energy reference surface. In contrary to Pauli's argument, all mentioned here do not resort to any additional assumption. As well known, the signs of $\bar{x}$ and $\bar{p}$ depend on their directions, while the signs of $t$ and $p_0$ depend on the choice of zero-reference-point. Meanwhile, the related observable quantities depend only on the difference, not on their absolute values.

If $\psi^+(x)\psi(x)d\sigma^0$ is the probability of finding a particle in the spatial volume element $d\sigma^0$ at time $t$, then, in our framework, let $\int_{-\infty}^{+\infty}\psi^+(x)\psi(x)d\sigma^l = 1$ ($l = 0,1,2,3$), $\psi^+(x)\psi(x)d\sigma^l$ is the probability of finding a particle in the 3D hypersurface element $d\sigma^l$ at coordinate $x^l$. Similarly, we can define $\langle t \rangle \equiv \int_{-\infty}^{+\infty}\psi^+(x)t\psi(x)dt$.

In conclusion, it is reasonable to take $x^0 = t$ as time operator conjugating to $i\partial_0$ rather than $\hat{H}$. As a consequence, the Pauli's theorem no longer holds true in this case. Although it is the most appropriate choice to take $x^0 = t$ as a parameter (because of: (1). this moment the law of causality is naturally preserved; (2). time is one-dimensional while space three-dimensional), $x^0 = t$ also plays the role of time operator in quantum mechanics and quantum field theory. In other words, the reasons for choosing time as parameter lie in not so much ontology as methodology and epistemology.

## Appendix A

### Time Operator is Time Dependent

In fact, for the 4D angular momentum tensor operator $J^{\mu\nu}$ of a free Dirac field, we have

$$\hat{H} = \vec{\alpha} \cdot \hat{p} + \beta m, \quad J^{\mu\nu} = ix^\mu \partial^\nu - ix^\nu \partial^\mu + S^{\mu\nu} \tag{A-1}$$

$$\frac{dJ^{\mu\nu}}{dt} = \frac{\partial J^{\mu\nu}}{\partial t} + i[\hat{H}, J^{\mu\nu}] = 0 \tag{A-2}$$

where $S^{\mu\nu} = \frac{i}{4}[\gamma^\mu, \gamma^\nu]$ is the 4D spin tensor (μ, ν = 0, 1, 2, 3), $\gamma^\mu, \vec{\alpha} = \gamma^0 \vec{\gamma}$ and $\beta = \gamma^0$ are the Dirac matrices. Obviously, $x^0 = t$ presenting in the operator $J^{\mu\nu}$ plays the role of time operator, and then we have $\hat{T} = x^0$. In view of the fact that $[\hat{H}, i\partial^\mu] = 0$, $[\hat{H}, \vec{x}] = -i\vec{\alpha}$ and $\hat{H}\vec{\alpha} + \vec{\alpha}\hat{H} = -2i\vec{\partial}$, by using (A-2) (let $_\mu$ or $\nu = 0$), we have $[\hat{H}, x^0] = 0$, which can also be directly derived from (A-1). Therefore, (A-2) implies that $\frac{d\hat{T}}{dt} = \frac{\partial \hat{T}}{\partial t} + i[\hat{H}, \hat{T}] = \frac{\partial \hat{T}}{\partial t}$.

## Appendix B

### 4D Angular Momentum Tensor of a Charged Field

In an analogous manner (see for example, J. D. Jackson, *Classical Electrodynamics* (by John Wiley & Sons. Inc., 1975), P.180), we can extend the traditional relation between 3D angular momentum and magnetic moment to the 4D tensor case. As for the electromagnetic vector potential $A_\mu(x)$, we have

$$A^\mu(\bar{x}, t) = \frac{d}{dt'} \int A^\mu(\bar{x}, t) dt' = \frac{1}{4\pi} \int \frac{J^\mu(\bar{x}', t')}{|\bar{x} - \bar{x}'|} d^3 x'$$

$$= \frac{1}{|\bar{x}|} \int J^\mu(\bar{x}', t') d^3 x' + \frac{\bar{x}}{|\bar{x}|^3} \int J^\mu(\bar{x}', t') \bar{x}' d^3 x' + \cdots \cdots$$

(B-1)

where $t' = t - \frac{r}{c}, r = |\bar{x} - \bar{x}'|$, $J^\mu$ is a localized divergenceless current, which permits simplification and transformation of the expansion (B-1). Let $f(x')$ and $g(x')$ be well-behaved functions of $x'$ to be chosen below

$$\int (fJ \cdot \Delta g + gJ \cdot \Delta f) d^4 x = 0 \qquad (\Delta \cdot J = 0)$$

(B-2)

where $\Delta$ denotes the 4D gradient operator. Let $f = x_\mu$ and $g = x_\nu$, we have

$$\int (x_\mu J_\nu + x_\nu J_\mu) = 0$$

(B-3)

$$\bar{x} \frac{d}{dt'} \int J_\mu(x') \bar{x}' d^4 x' = \sum_j x_j \frac{d}{dt'} \int x'_j J_\mu(x') d^4 x'$$

$$= -\frac{1}{2} \sum_j x_j \frac{d}{dt'} \int (x'_i J_\mu - x'_\mu J_j) d^4 x'$$

(B-4)

It is customary to define the define the electromagnetic moment density

$$m^{\mu\nu} = \frac{1}{2} [x^\mu J^\nu - x^\nu J^\mu]$$

(B-5)

and its integral as the electromagnetic moment

$$M^{\mu\nu} = \frac{1}{2} \int [x'^\mu J^\nu - x'^\nu J^\mu] d^3 x'$$

(B-6)

Assuming that $J^\mu$ is provided by $N$ charged particles with momentums $p_n^\mu = m_0 u_n^\mu$ (n = 1,2,..., N) and charges $e$, then $J^\mu(x') = \sum_n e u_n^\mu(t') \delta^3(\bar{x} - \bar{x}'_n) \frac{d\tau}{d\tau'}$, where $\tau, \tau'$ are the proper times. When $t = t'$, we have

$$M^{\mu\nu} = \frac{e}{2m_0} \sum_n (x_n^\mu p_n^\nu - x_n^\nu p_n^\mu) \frac{d\tau}{dt} = \frac{e}{2m} L^{\mu\nu}$$

(B-7)

where $m$ is the relativistic mass, $L^{\mu\nu}$ is the 4D angular momentum tensor.

## Appendix C

## Another Origin of Eq.23

In fact, Eq.23 can be obtained in another manner: we impose the boundary conditions

$$\phi(x), \partial_\mu \phi(x) \to 0, \text{ as } x^\mu \to \pm\infty, \mu = 0,1,2,3 \tag{C-1}$$

on the field $\phi(x)$, then an equation of continuity $\frac{\partial j_\mu(x)}{\partial x_\mu} = \partial^\mu j_\mu = 0$ associated with Noether's theorem can be integrated over 3D hypersurface and theorem of Gauss be used:

$$0 = \int_{\sigma^l} \partial^\mu j_\mu d\sigma^l = \int_{\sigma^l} \partial^l j_l d\sigma^l + \oint_{\partial\sigma^l} j_m dS^m$$

(C-2)

Where $m, l = 0,1,2,3$ and $m \neq l$. The value of the integral over the 2D surface $\partial\sigma^l$ vanishes since the fields and their derivatives are assumed to fall off sufficiently. Therefore

$$\partial^l \int_{\sigma^l} j_l d\sigma^l = 0 \text{ (the index } l \text{ is not summed)} \tag{C-3}$$

Namely, $G_l \equiv \int_{\sigma^l} j_l d\sigma^l$ is a quantity being independent of $x^l$ (we call $G_l$ as *generalized conserved quantity with respect to $x^l$*). Now, let the current density $j_\mu = T_{\mu\nu}$, where $T_{\mu\nu}$ is the canonical energy-momentum tensor of a field:

$$T_{\mu\nu} = \frac{\partial \Gamma}{\partial \partial^\mu \phi} \partial_\nu \phi - g_{\mu\nu} \Gamma \tag{C-4}$$

Obviously, the corresponding generalized conserved quantity with respect to $x^l$ is the generalized Hamiltonian $H_l$:

$$G_l = \int d\sigma^l T_{ll} = H_l \tag{C-5}$$

, which is exactly the same as Eq.23.